\documentclass[twocolumn,showpacs,preprintnumbers,amsmath,amssymb]{revtex4}

\usepackage{graphicx}
\usepackage{dcolumn}
\usepackage{bm}
\usepackage{amssymb}

\def\address{\@ifstar{\address@star}%
  {\@ifnextchar[{\address@optarg}{\address@noptarg}}}

\begin{document}

\author{S.N.~Gninenko}
\author{N.V.~Krasnikov}
\author{V.A.~Matveev}

\affiliation{Institute for Nuclear Research of the Russian Academy of Sciences, Moscow 117312}


\title{Invisible $Z'$ as a probe of extra dimensions at the CERN LHC}

\date{\today}

\begin{abstract} 
A class of extra dimensional models with a warped metric predict 
tunneling of a massive particle
localized on our brane and escaping  into additional  dimensions. 
The experimental signature of this effect is the disappearance of the  particle
 from our world, i.e. the $particle \to invisible$ decay.
We  point out that measurements of $Z'\to invisible$ decay width 
of a new heavy gauge boson $Z'$ at the CERN LHC can be 
effectively used  to probe the existence of large extra dimensions.
This result enhances motivations for a 
more sensitive search and study  for this decay mode and suggests 
additional direction for testing  extra dimensions in collider experiments. 
\end{abstract}

\maketitle


Presently there is a big interest in models with
additional dimensions \cite{rus1}-\cite{krasn} 
which might provide  solution to the 
gauge hierarchy problem \cite{rand1}-\cite{anton2},
 for a  review see e.g. \cite{rubakov}.
For instance, as it has been shown  
in the five dimensional model, so called RS 2-model \cite{rand2},  
there exists a thin-brane solution to the 5-dimensional 
Einstein equations which has flat 4-dimensional hypersurfaces,
\begin{equation}
ds^2 = a^2(z)\eta_{\mu\nu}dx^{\mu}dx^{\nu} -dz^2.
\end{equation} 
Here
\begin{equation}
a(z) = exp(-k|z|)
\end{equation}
and the parameter $k > 0 $ is determined by the 5-dimensional Planck 
mass and bulk cosmological constant and  represents a scale of new physics.

Not long ago, a peculiar feature of massive matter in brane world has been 
reported \cite{tinyakov}. It has been shown that particles initially located 
on our brane may leave the brane and disappear into extra dimensions.
 These kind of transitions have been found to be generic in a class of models
of localization of particles on a brane. The localization becomes incomplete 
if particles get masses and they could tunnel from the brane into extra 
dimensions. The experimental signature of this
effect is the disappearance of a particle in our world,
i.e. the $particle \to invisible$ decay. 

Particles disappearing into bulk are inherent also in 
string theory, e.g.  the disappearance into bulk have been
found in the context of noncommutative gauge theories \cite{drs}. 
The difference, though, is that massless gauge bosons (photon and gluon) as well
 as particles charged under unbroken gauge group  are strictly bound to the 
brane, and only massive and
electrically neutral gauge bosons can disappear into bulk. Interestingly,  
this process in the noncommutative soliton
context is opened up by the Higgs mechanism \cite{drs}. For disappearance
processes to be reasonably fast, the string energy scale should be in 1-10 TeV
range. Such low string scale is behind
many ideas beyond the SM, see, e.g.,  Ref. \cite{drs} and references therein.

 The case of the electromagnetic field propagating in 
the Randall-Sundram type of metric in the presence of extra compact 
dimensions \cite{oda,rub} has been considered in \cite{rub}, where it
 was shown that 
 the transition rate  of a virtual photon into additional dimensions is 
different from zero. This effect 
 could result in disappearance of  a neutral system,
orthopositronium ($o-Ps$) \cite{gkr}, a triplet bound state of an electron
and positron,   at a rate within
 two orders of magnitude of the present best experimental limit
on the  branching ratio of the $o-Ps \to invisible$ decay
$Br(o-Ps \to inv) < 4.3 \times 10^{-7}$ (90\% C.L.)
from the recent ETH-INR experiment \cite{bader}.

The lower limit on the $k-$parameter, which can be extracted from the results of this experiment is  
\begin{equation}
k\gtrsim 0.5 TeV
\label{ops}
\end{equation} 
Stronger bounds on the parameter $k$ for photons escaping into extra 
dimensions  can be obtained  from astrophysical considerations \cite{fg}.

\begin{figure}[h]
\begin{center}
    \resizebox{4cm}{!}{\includegraphics{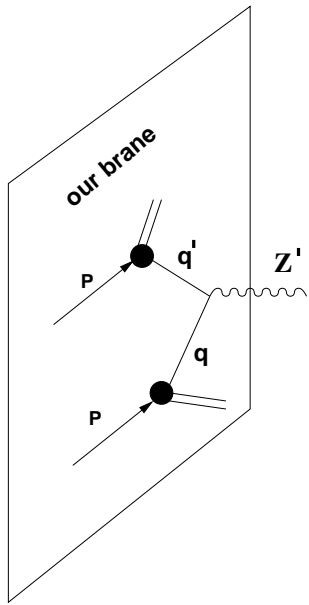}}
     \caption{Schematic illustration of the production and disappearance 
processes of  $Z'$ bosons in $pp$ collisions at LHC: protons and
quarks propagate along the our brane while $Z'$ escapes into extra dimensions. }
\label{zedm}
\end{center}
\end{figure}

Consider now the disappearance of the  $Z$ bosons. 
The transition rate into additional dimension(s) of the  $Z$ on-shell   
 is given by \cite{rub,gkr}  
\begin{equation}
\Gamma = q(n) M_Z \bigl(\frac{M_Z}{k}\bigr)^n
\label{z1}
\end{equation}
where $q(n)$ is a numerical coefficient, and $M_Z$ is the $Z$ mass.
To make a quantitative estimate, we take 
$n=2$ ( (4+2+1)-dimensional space-time). In this case the $Z$ disappearance 
rate is given by \cite{gkr}
\begin{equation}
\Gamma = \frac{\pi M_Z}{16} \bigl(\frac{M_Z}{k}\bigr)^2
\label{z2}
\end{equation}
Important bounds on the parameter $k$ and 
$\Gamma(Z \rightarrow add.~dim.)$ arise from the combined LEP result
 on the precise measurements of the total and partial $Z$ 
widths \cite{gkr,pdg}:
\begin{equation}
k \gtrsim 17~TeV 
\label{lep1}
\end{equation}
The limit of Eq.(\ref{lep1}) is much stronger than the one of Eq.(\ref{ops})
obtained from positronium  measurements. 
Note that combined result on direct LEP measurements of the invisible width \cite{lep} gives less stringent limit \cite{gkr}:  
\begin{eqnarray}
k \gtrsim 4~TeV 
\label{lep2}
\end{eqnarray}
To solve the gauge hierarchy 
problem models with additional dimension(s) one may expect
\begin{equation}
k \simeq 10~TeV
\label{hie}
\end{equation}
which is consistent with bounds of Eqs.(\ref{ops},\ref{lep1},\ref{lep2}).

Similarly, consider now the new TeV-scale boson $Z'$, which 
 appears in many models of physics beyond the SM, 
for a review see e.g. \cite{review}.   $Z'$ bosons that decay to leptons
have a simple, clean experimental signature, and consequently can be searched for up to high masses at colliders. Current direct search limits from Tevatron experiments restrict the $Z'$ mass to be greater than about 900 GeV when its couplings to SM fermions are identical to  those of the Z boson \cite{pdg}. 

The $Z'$  is  a good candidate 
for the searching for effect of disappearance into 
additional dimension(s), 
since it potentially could be discovered at the LHC with a mass 
 up to 5 TeV, see e.g.
\cite{cmstdr}. Schematically, the disappearance
of $Z'$ produced in $pp$ collisions into extra dimensions is illustrated in 
Fig. \ref{zedm}. As it follows from Eqs.(\ref{z2},\ref{hie}), for 
$M_Z\lesssim 5$ TeV  one may expect 
\begin{equation}
\Gamma (Z' \rightarrow add~dim) \lesssim 10~GeV
\end{equation}

Consider, for comparison   
the $Z'\to \nu \nu$ decay rate to the SM neutrinos  
 in several canonical $Z'$ models \cite{review}. First we shortly 
describe these models and the SM fermions couplings the $Z'$.

\begin{itemize}

\item the E$_6$ models  - are described by the breaking chain
$E_6 \to SO(10)\times U(1)_\psi \to SU(5)\times U(1)_\chi \times U(1)_\psi
\to SM\times U(1)_\beta $. Many studies of $Z'$ are focusing on the two 
extra $U(1)'$ which occur in the above decomposition of the $E_6$.
The lightest $Z'$ is defined as :
\begin{equation}
Z'= Z'_\chi cos\beta  + Z'_\psi sin\beta 
\label{z'}
\end{equation} 
where the values $\beta = 0$ and $\beta= \pi/2$ corresponds to 
pure $Z'_\chi$ and $Z'_\psi$ states of the $\chi$- and $\psi$-model,
 respectively. 
The value $\beta= arctan(-\sqrt{5/3})$
is related to a $Z'_\eta$ boson that would originate from the direct breaking 
of $E_6$ to a rank-5 group in superstrings inspired models.

\item  the  Left-Right
Symmetric (LRSM) model  is based on the symmetry group 
$SU_C(3) \otimes SU_L(2) \otimes SU_R(2)\otimes U(1)_{B-L}$ \cite{lr}, 
in which $B$ and $L$ are the baryon and lepton numbers, respectively.
The model necessarily incorporates  
three additional gauge bosons $W^\pm_R$ and $Z'$.
The most general $Z'$ is coupled  to a linear combination of right-handed 
and $B-L$ currents:
\begin{equation}
J^\mu_{LR}=\alpha_{LR}J^\mu_{3R}-(1/2\alpha_{LR})J^\mu_{B-L}
\end{equation}
where $\alpha_{LR}=\sqrt{(c_W^2g_R^2)/s_W^2g_L^2)-1}$, with 
$g_L=e/s_W$ and $g_R$ are the $SU(2)_L$ and $SU(2)_R$ coupling
constant with $s_W^2=1-c_W^2=sin^2 \Theta_W$. The $\alpha_{LR}$-parameter 
is restricted to be in the range $\sqrt{2/3} \lesssim \alpha_{LR} \lesssim \sqrt{2}$. The upper bound corresponds to the so-called manifest LRSM with $g_l=g_R$,
 while 
the lower bound corresponds to the $\chi$-model discussed above, since 
$SO(10)$ can results to both $SU(5) \times U(1)$ and $SU(2)_R \otimes SU(2)_L \otimes U(1)$ breaking parameter. 

\item in the sequential model (SSM) the corresponding $Z'$ boson has the 
same couplings to fermions as the $Z$ of the SM. The $Z'$ could be 
considered as an excited state of the ordinary $Z$ in models with 
extra dimensions at the weak scale.    

\end{itemize}

The $Z'$ boson partial decay width into a fermion-antifermion pair 
 is given by 
\begin{eqnarray}
&&\Gamma(Z' \to f\overline{f})=N_C\frac{\alpha M_{Z'}}{6c_W^2}\sqrt{1-4\eta_f}\times \nonumber\\
&&[ (1+2\eta_f)(g^{f}_L)^2+(1-4\eta_f)(g^{f}_R)^2 ]
\end{eqnarray}
where $N_C$ is a color factor ($N_C=3$ for quarks and $N_C=1$ for leptons),
$g^{f}_L, g^{f}_R$ are the left- and right-handed couplings of the $Z'$ to 
the SM fermions,  $\alpha$ is the electromagnetic 
coupling constant, which is $\alpha \simeq 1/128$ at the $M_{Z'}$  scale, 
 and  $\sqrt{\eta_f}(=m_f/m_{Z`})$ is assumed to be  $\ll 1$.
The left-handed couplings of the $Z'$ to the SM neutrinos are 
$g^{f}_L = \frac{3cos \beta}{2\sqrt{6}}+\frac{\sqrt{10}sin\beta}{12}$ 
and $g^{f}_L = \frac{1}{2\alpha_{LR}}$ for E$_6$ and LRSM models, respectively,
while the right-handed couplings $g^{f}_R=0$ in both models.  
The $(g^{f}_L)^2$ is restricted to lie 
 in the range   
$0.07 (\beta\simeq \pi/2) \lesssim (g^{f}_L)^2 \lesssim 0.45~(\beta \simeq 0.4)$ for the
E$_6$ model\footnote{The minimal possible  value for  $(g^{f}_L)^2$
in this model is zero , we take it to be $\simeq 0.07$ for $\beta=\pi/2$
 to have non-zero $Z'-\nu$ coupling.} and in the range  $ 1/8 \lesssim (g^{f}_L)^2 \lesssim 3/8$
for the LRSM model.

\begin{figure}[tbh!]
\begin{center}
    \resizebox{6cm}{!}{\includegraphics{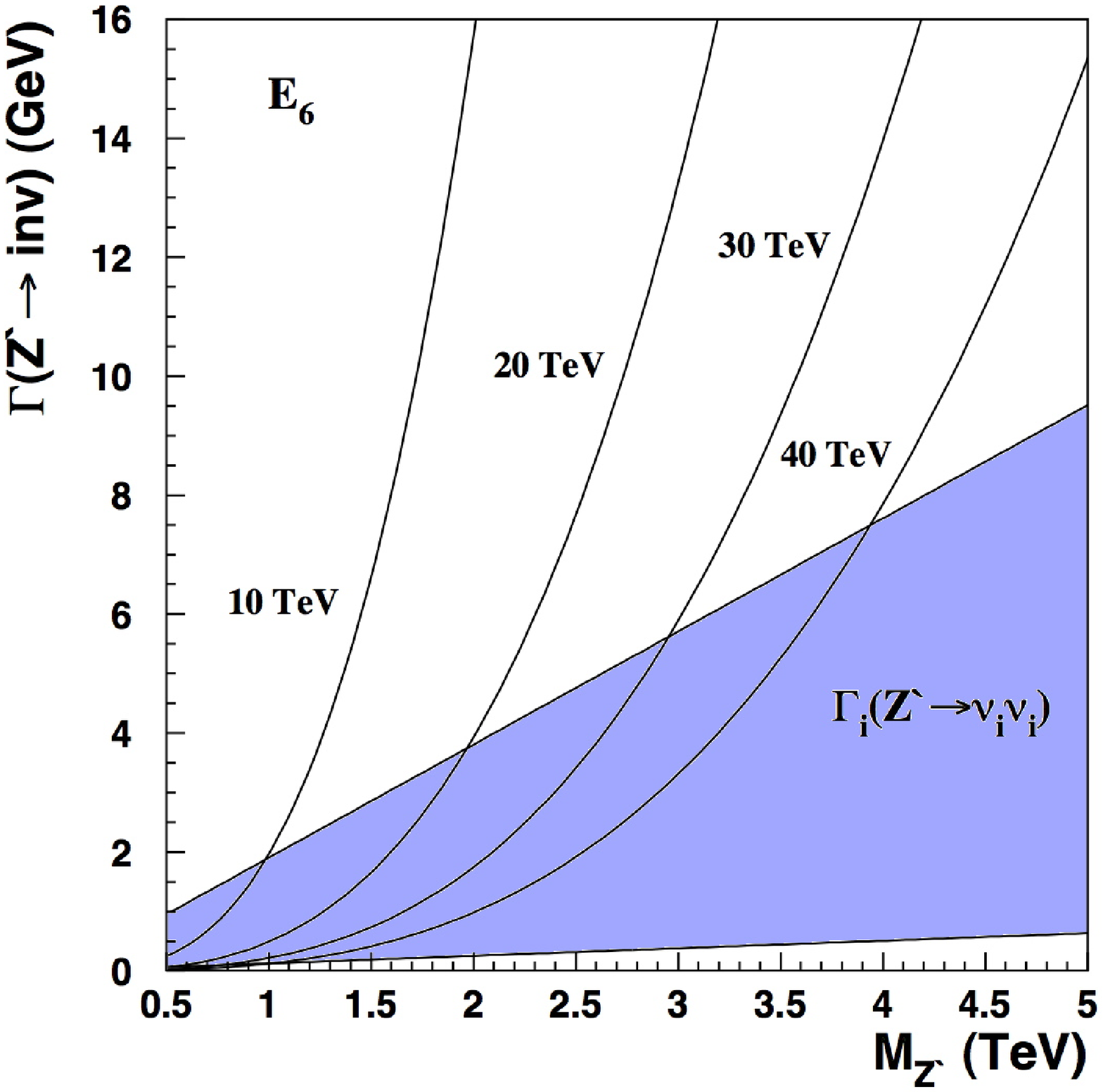}}
    \resizebox{6cm}{!}{\includegraphics{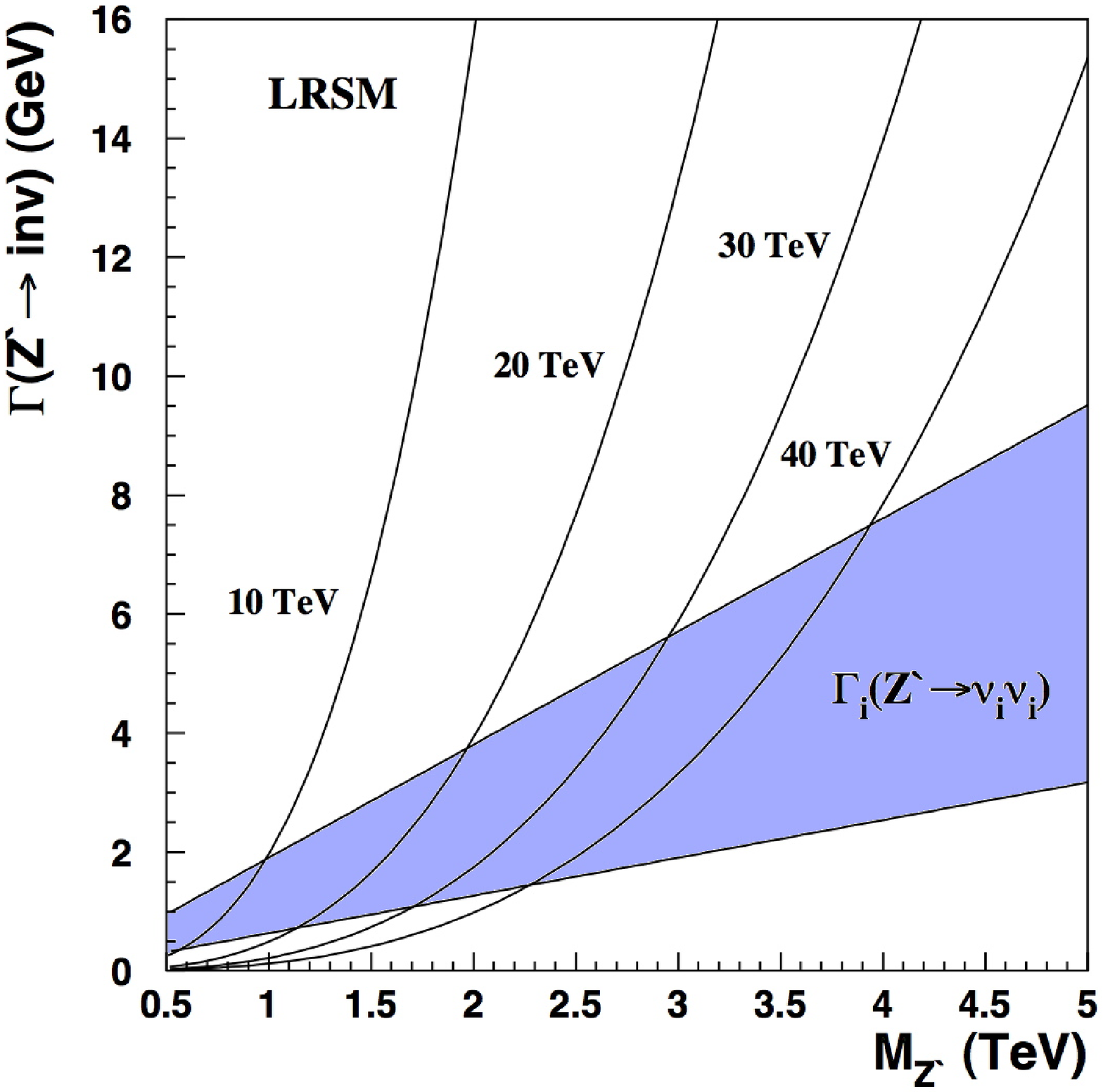}}
     \caption{ The dependence of the rate $\Gamma(Z'\to add~ dim)$
 as a function of the $Z'$ mass for different $k-$parameters shown 
near the curves. The shaded 
region  represents the possible values of  
 $ \Gamma_i(Z'\to \nu_i \overline{\nu}_i)$
decay rate calculated for E$_6$ (upper plot) and LRSM models taking into account 
uncertainties in the $g^{f}_L$ coupling (see text). }
\label{width}
\end{center}
\end{figure}

In  Fig.\ref{width}, 
the dependence of the rate $\Gamma(Z'\to add~ dim)$ and the total decay 
rate  $\Gamma_i(Z'\to \nu_i \overline{\nu}_i)$ of $Z'$ to the  
 SM neutrinos are shown as a function of the $Z'$ mass  for 
different $k-$ parameters. The results are obtained 
 for E$_6$ and LRSM models taking into account 
uncertainties in the $g_L^f$ coupling. 
 One can see, that 
for the most of the parameters space 
the $Z'\to add~dim$ rate  either is comparable 
or dominates the total  $\Gamma_i(Z'\to \nu_i \overline{\nu}_i)$ rate.
If, for example  the $Z'$ mass is $M_{Z'}= 1.5 $ TeV and $k\simeq 10$ TeV, the 
$\Gamma(Z'\to add~ dim)$  rate is about factor  2 greater  then 
the largest $\Gamma_i( Z'\to \nu_i \overline{\nu}_i)$ decay rate in the 
 LRSM model. Hence, 
 if $Z'$ is observed at the LHC,  the question  of accurate 
measurements of its properties and, in particular of its 
invisible decay rate, is  of
 great interest for possible  observation of extra dimensions. 

The experimental signature of $Z'\to inv$ process at the LHC is  
 the large missing $E_T~(\gtrsim 150- 200)$ GeV). This signature is relatively 
clean, however, in order to discover  process $Z'\to add~dim$   
 one has to distinguish whether the  extra invisible width $\Gamma (Z'\to add~dim)$ could 
be determined over the background from the 
 $Z'$ decays into SM neutrinos.
 
\begin{figure}[h]
\begin{center}
    \resizebox{6cm}{!}{\includegraphics{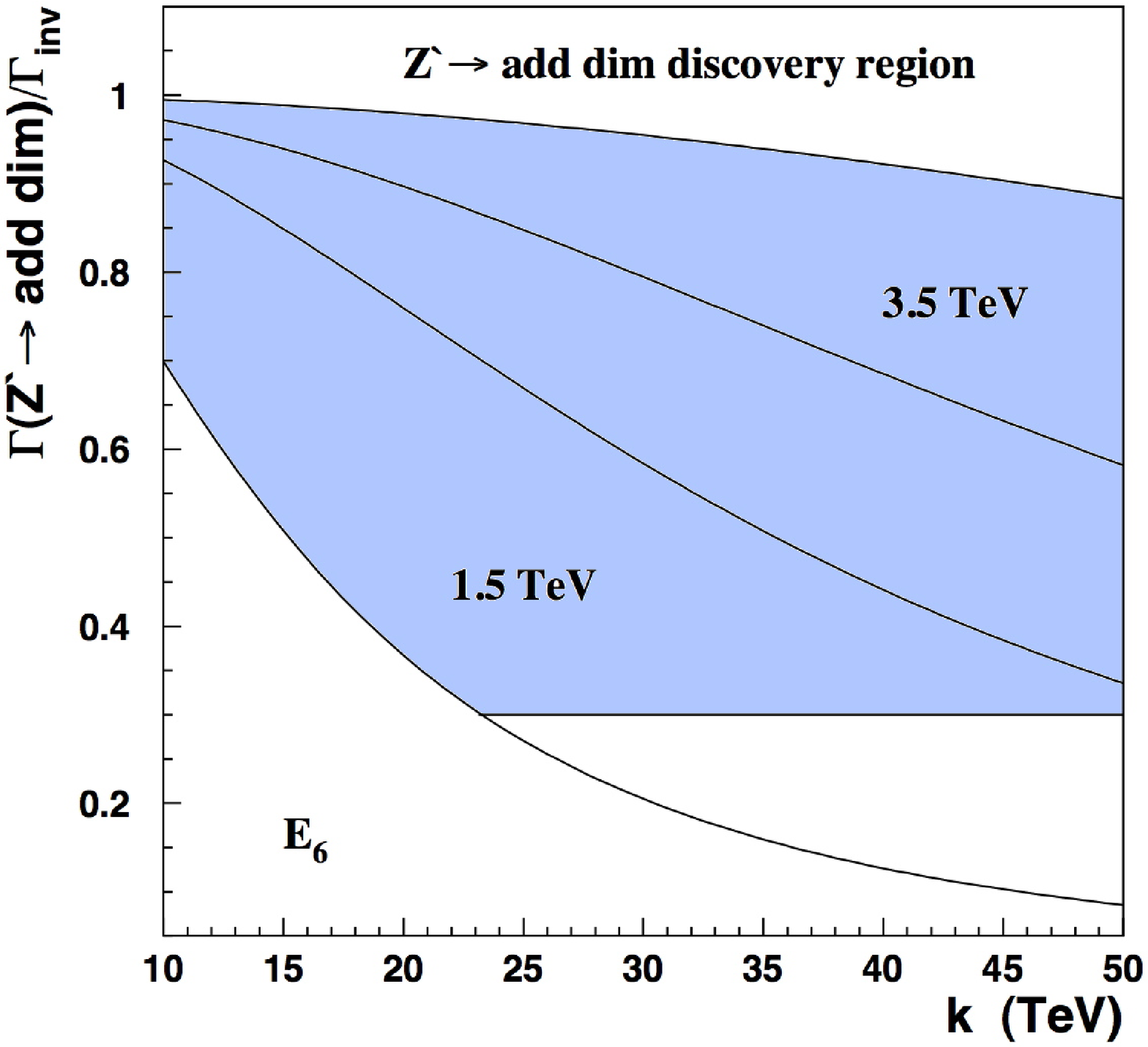}}
    \resizebox{6cm}{!}{\includegraphics{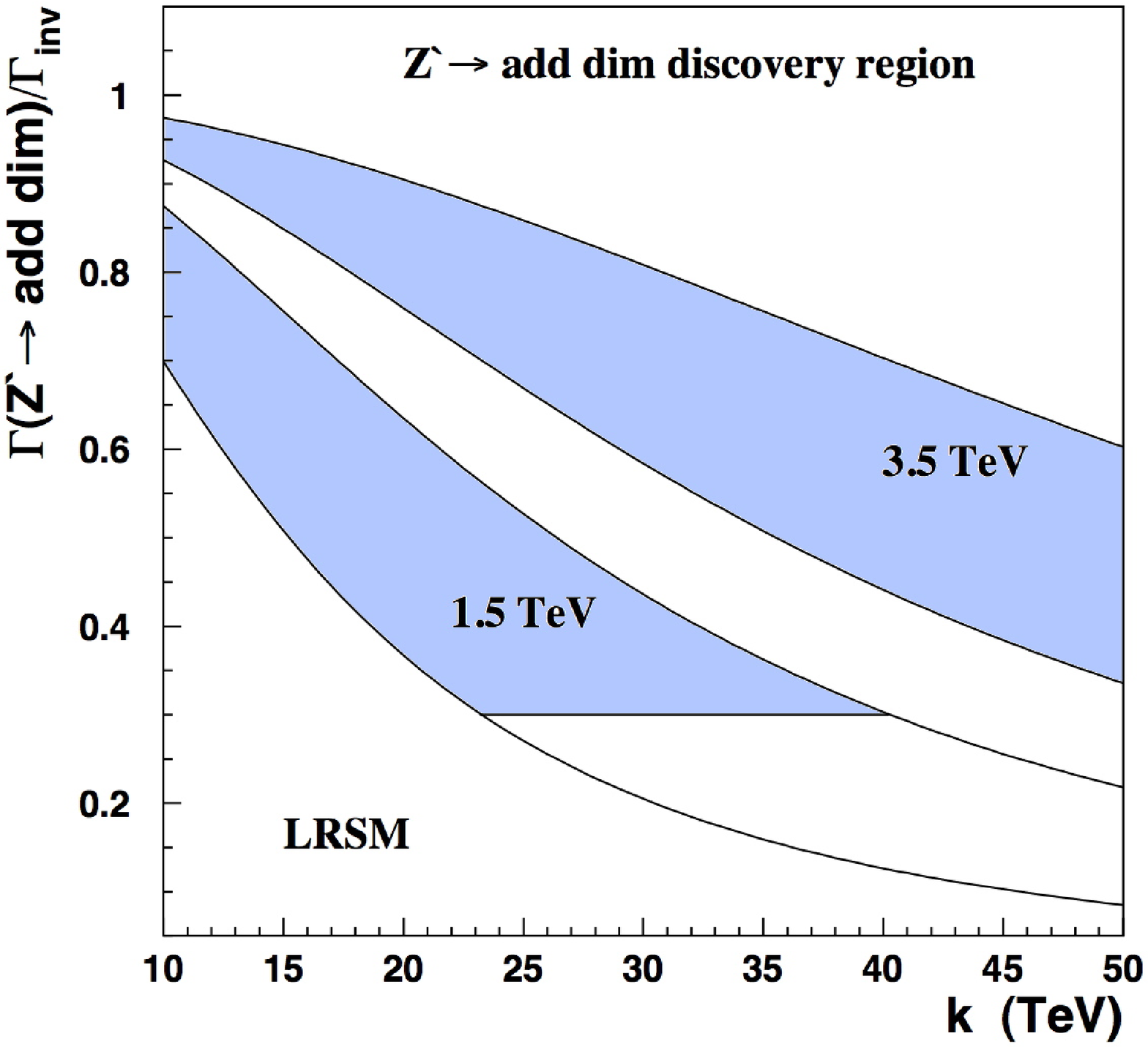}}
     \caption{  Two regions of possible
 branching fraction $\Gamma (Z'\to add~dim)/ \Gamma_{inv}$
in  E$_6$ (upper plot) and LRSM  models determined for  $M_{Z'}=1.5$ and 3.5 TeV by   
 taking into account uncertainties in the $g^{f}_L$ coupling.
The central part in the upper plot is  the overlap between the regions, see
plot below for comparison. The shaded areas corresponds to 
$\Gamma (Z'\to add~dim)/ \Gamma_{inv}\gtrsim 0.3$ for which the  process 
$Z'\to add~dim$  could  be seen  at the LHC for the integrated 
luminosity of 30 fb$^{-1}$. }
\label{ratio}
\end{center}
\end{figure}

Recently, an interesting analysis relevant to the present discussion
 has been reported  \cite{mc,pq,pqz}. 
It has been shown that among several reactions of  $Z'$ production 
in $pp$ collisions at the LHC, the reaction $pp \to ZZ'\to l^+ l^- + 
E_T^{miss}$ of associated $Z$ and $Z'$ production 
  is quite convenient to study $Z'\to inv$ decay properties. 
It has been  demonstrated that for the most popular models the 
 invisible $Z'$ decay can be seen over the 
SM background with a significance of $S/\sqrt{B} = 3$  at 10 fb$^{-1}$, while 
$S/\sqrt{B} = 5$  can be reached with 30 fb$^{-1}$.

An important observation is that if the only 
invisible decays of the $Z'$ are to SM neutrinos, the invisible width 
of this process can be predicted from the analysis of 
the Drell-Yan $Z'$ production \cite{pq}.
Then, the additional contribution $\Delta \Gamma_{inv}$ to the 
$\Gamma_{inv}$ can be determined as an excess over expected 
$pp \to ZZ'\to l^+ l^- + E_T^{miss}$
cross-section predicted by the analysis of on-peak data.
The initial analysis demonstrates that if $\Delta \Gamma_{inv}$ contributes 
to the total invisible width at the level of $\gtrsim 30$\%, the corresponding 
underlying process can be discovered at the LHC for the
integrated luminosity of 30 fb$^{-1}$ \cite{pqz}.

In Fig.\ref{ratio}, the  
regions of possible values for the decay branching fraction 
$R=\Gamma(Z'\to add~ dim)/\Gamma_{inv}$ ( here $\Gamma_{inv}=\Gamma(Z'\to add~ dim) + \Gamma_i(Z'\to \nu_i \overline{\nu}_i)$)  
calculated in  E$_6$ and LRSM models, respectively, 
are shown for  $Z'$ masses of 1.5 and 3.5 TeV. 
One can see that the significant part 
of  the $(R,k)$ parameter space satisfies the condition
 $R\gtrsim 0.3$ even with  big uncertainties in the $g^{f}_L$ 
coupling, thus making the process $Z'\to add~dim$ feasible for 
observation at the LHC.

In summary,  we  have demonstrated 
 that measurements of the $Z'\to inv$ decay suggest an 
interesting additional direction to probe  extra  dimensional physics at the
 CERN LHC. Although  the results presented  are model-dependent, 
we believe that they strengthen  current motivations 
and  justify efforts for more sensitive search for the $Z'\to inv$ decay 
and accurate measurements of its properties
in LHC experiments. 
Detail simulations work, which is beyond the scope of this paper,
is in progress \cite{sng}.
       

This work grew in part from our participation in 
the 12th RDMS CMS Conference on physics at LHC. We wish to thank 
organizers of this conference, in particular N.M. Shumeiko, for their 
warm hospitality in Minsk. We thank V.A. Rubakov for valuable discussions and 
 suggestions, and 
M.M. Kirsanov for help in simulations.
The work was supported by Grants RFFI 03-02-16933 and
RFFI 08-02-91007-CERNa.

\end{document}